\documentclass[aps,
prd,twocolumn,
prabib,
showpacs,nofootinbib,showkeys,10pt]{revtex4-2}
\usepackage[table]{xcolor}
\usepackage{graphicx} \usepackage{amsmath} \usepackage{amssymb}
\usepackage{amsfonts} \usepackage{bm}
\usepackage{array}
\usepackage{siunitx}
\usepackage[singlelinecheck=false]{caption}

\usepackage{ytableau}
\usepackage{multirow}
\usepackage{mathtools}
\usepackage{tikz}

\begin{document}

\newcommand{\be}{\begin{equation}} \newcommand{\ee}{\end{equation}}
\newcommand{\bea}{\begin{eqnarray}}\newcommand{\eea}{\end{eqnarray}}

\title{Multi-target  quantum walk search    on Johnson graph}

\author{Pulak Ranjan Giri} \email{pu-giri@kddi-research.jp}

\affiliation{KDDI Research,  Inc.,  Fujimino-shi, Saitama, Japan} 

\begin{abstract} 
The  discrete-time quantum walk  on  the Johnson graph  $J(n,k)$ is a  useful tool   for  performing   target  vertex searches    with  high success probability. 
This   graph is defined by  $n$  distinct elements, with vertices being all the   \(\binom{n}{k}\)  $k$-element subsets and two vertices are connected by an  edge  if they differ  exactly by  one element.  However, most  works in the literature   focus   solely on  the  search for a single target   vertex on the Johnson graph.  In this article,  we  utilize  lackadaisical quantum  walk--a form of discrete-time coined quantum walk  with a wighted self-loop at each vertex of the graph--along with our recently proposed modified coin operator,  $\mathcal{C}_g$,   to find multiple target   vertices on  the  Johnson graph $J(n,k)$ for various  values  of  $k$.  Additionally, a  comparison based on the numerical analysis   of  the performance of  the $\mathcal{C}_g$ coin operator  in searching  for multiple target  vertices  on the Johnson graph,   against   various other  frequently used  coin operators   by the  discrete-time quantum walk  search algorithms,   shows  that  only   $\mathcal{C}_g$  coin  can search  for  multiple target  vertices with a very  high  success probability in all   the scenarios discussed in this  article,  outperforming other  widely used coin operators  in the literature.  

 \end{abstract}


\date{\today}

\maketitle 



\section{Introduction} \label{in}
\begin{figure*}
  \centering
     \includegraphics[width=0.8\textwidth]{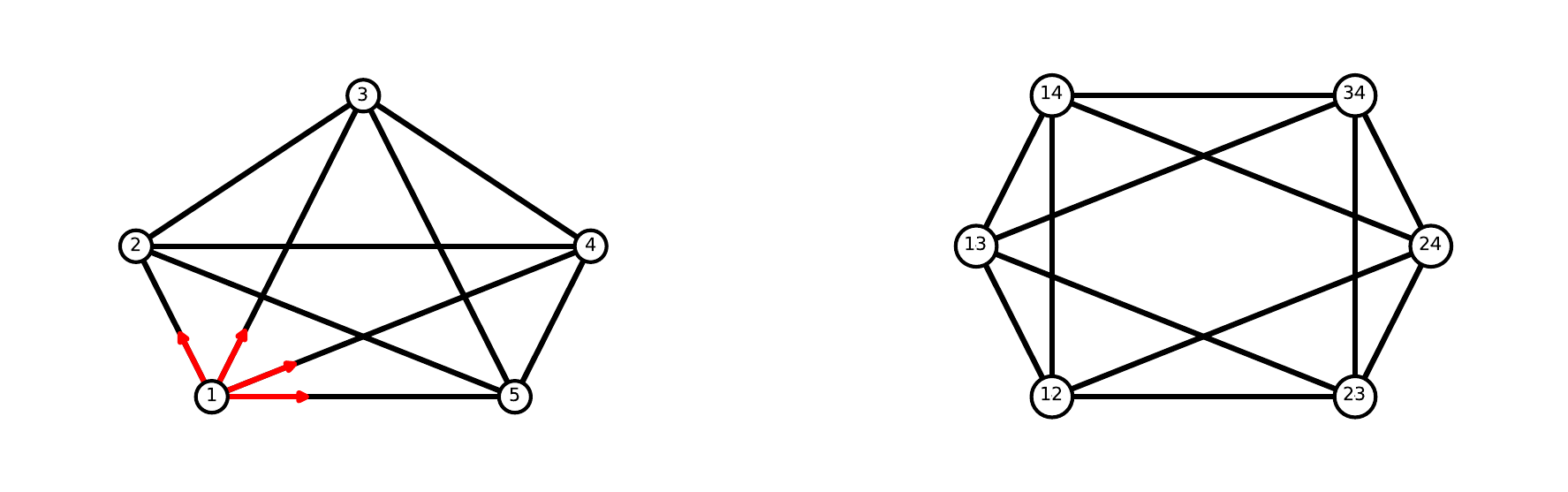}
          
       \caption{\textbf{Johnson graph $J(5,1)$(left)  and  $J(4,2)$(right).}}
\label{fig1}       
\end{figure*}

Johnson graph, $J(n,k)$,  \cite{godsil}  consists of   a set of  $n$ distinct  elements.  The vertices of this  graph are  formed by taking $k$ elements    from the set,  resulting in     
$N$= \(\binom{n}{k}\) vertices, each having  a degree of  $d= k(n-k)$.   Two vertices are considered  nearest neighbor and connected by an edge  if they differ  by  only   one element. Several  known  graphs can be obtained from the Johnson graph by varying  the parameter $k$.  For example, the complete graph  $K_N$   can be obtained from the Johnson graph when  $k=1$, where it has  $N = n$  vertices and  each vertex has $d= n-1$ edges.   In  fig.  \ref{fig1}(left) a  complete graph $K_5 = J(5,1)$  with $n=5$ elements  
$\{1, 2, 3, 4, 5\}$ is presented. In this case,  each element corresponds to  a vertex  of the graph.   Another example is the triangular graph, $T_n$,  which  can be obtained from the Johnson graph when  $k=2$.  
In  fig.  \ref{fig1}(right) a  triangular  graph $T_4 = J(4,2)$  with $n = 4$ elements  $\{1, 2, 3, 4\}$ is presented. In this case, the   six vertices  $\{12, 13, 14, 23, 24, 34 \}$  are the $2$-element subsets of the elements  $\{1, 2, 3, 4\}$.

Johnson graph $J(n,k)$  is  an  interesting graph  structure for  the  quantum computing community, particularly   for the  study of  quantum walk \cite{portugal}.   It is the only graph that prevents  the quasi-polynomial  algorithm  \cite{babai} for  graph isomorphism   from being polynomial.  The Johnson graph   has been used to study the element distinctness algorithm  in  both discrete-time  \cite{amelement} and continuous-time  \cite{chelement} quantum walk.  Several studies on single target searches  on  Johnson graphs,  $J(n,k)$, using  both  continuous-time \cite{wongjpa} and discrete-time \cite{rho,rapo,tana,peng} quantum walks, have demonstrated  that it is possible to conduct searches   in optimal time with very  high success probability.  

In this paper,   we  study  multi-target spatial search   on the  Johnson graph  using the  discrete-time version of the  quantum walk (QW) with   our recently proposed  coin $\mathcal{C}_g$. We  compare the results  with  the performance of  other frequently used  coins in the literature.    It is noteworthy  that  the  multi-target spatial search, which we study in this article,  has applications  in  image processing \cite{giripla} and  the  Johnson graph  plays a crucial role   in generating secure hash functions \cite{cao}.

Quantum walk--a quantum counterpart of classical random walk--is a universal tool for  quantum computation \cite{childsprl}. It has been used in several quantum algorithms, such as spatial search \cite{portugal},  element distinctness \cite{amelement,chelement}, solving boolean formulas \cite{childstc}  and also in several  applications, such as  quantum hash function \cite{li}, and quantum edge detection \cite{giripla}.  
Both  continuous- \cite{childs} and   discrete-time \cite{amba2} versions of the   quantum walk   can     be    used to perform spatial search   on a  variety of different graphs.
However, this  generalization of the  celebrated Grover search \cite{grover1} to the spatial search on graphs  in the form of a quantum walk was not that  straightforward initially.  This is because,   in spatial search on graphs we are only allowed to shift from a vertex to the next nearest neighbor vertices  at a time, where a vertex may not be a nearest neighbor  to all the other vertices on the graph.  It was argued by Paul Benioff  \cite{beni}  that the  quantum search on a  graph with $N$ vertices will lose the   quantum speedup, because,   both the   iterations and reflection  need    $\mathcal{O}(\sqrt{N})$ time each,  making the total time  complexity no better than the time for a classical  exhaustive search on  an unsorted database.  However,  the claim  of   Paul Benioff   was   later  refuted  in ref.  \cite{amba1} by showing that it is possible to  do spatial search on a  graph faster than a classical exhaustive search. 

Note that,  a classical computer takes  $\mathcal{O}(N)$ time to search for a single target from an unsorted database of size $N$.  Grover's original  search  and its generalizations \cite{giri}  are  quadratically faster \cite{ni} than the classical exhaustive search.  The same  quadratic speedup  can also be achieved for spatial search  by a  quantum walk   on several   graphs \cite{she,meyer,amba4}.  

For the  quantum computer  to be efficient,  it is very important to have a high enough  output probability of the desired result as well as fast computation time   in order to accurately   measure it  in less time compared to its  classical counterparts.  Quantum walk search on graphs sometimes suffers  on  both counts, specifically, while searching  for multiple target vertices  on  graphs.  In general, quantum walk search  on  one- and two-dimensional square lattices  cannot  achieve  the    optimal speed  of $\mathcal{O}(\sqrt{N})$ for  quantum search.   Although, time complexity of  $\mathcal{O}(N)$  \cite{girimpa}   for a   one-dimensional  periodic  lattice is still quadratically faster than its corresponding time complexity for a  classical random walk.  Usually, in a  two-dimensional  periodic   lattice,    time complexity   is  $\sqrt{\log N}$ times greater  than the optimal time for quantum search.  However,  in a two dimensional lattice,   optimal speed   can be  achieved  in continuous-  \cite{tomo} and discrete-time  \cite{giri3}   quantum walk using extra long-range edges, such as the Hanoi network \cite{giriijqi} of degree four, HN4.    

Searching  for multiple targets on  graphs using a  quantum walk is usually challenging in terms of the  success probability of the target states and execution  time.  For example,  a pair of adjacent  target vertices \cite{rivosh1,rivosh,nah1} on a two-dimensional periodic lattice cannot  be found by the quantum walk with the  Grover coin, $\mathcal{C}_{grov}$ due to the existence of stationary states \cite{wong4}.  More generally,   exceptional  configurations of target vertices on a two-dimensional grid, which are of the  form  $2k \times m$ or  $k \times 2m$, \cite{men} for any positive $k, m$ also cannot be found. Even a lackadaisical quantum walk \cite{wong1,nahsof} with the  Grover coin,  $\mathcal{C}_{l}$, \cite{giriijqi24} cannot find these configurations.   Although,   target vertices of the form $k \times m$, for both  $k,m$  being odd, can be  found by a quantum walk search with the  Grover coin.  Similarly,  the SKW coin, $\mathcal{C}_{skw}$,  cannot  find  target vertices arranged along the diagonal  \cite{giriijqi24,amba5} of a two-dimensional square lattice.  Also configurations obtained by shifting  and/or rotating the diagonal configurations by $\pi/2$  \cite{men} cannot be found by the SKW coin.   Sometimes, though the success probability  is  high,    the running time increases when  searching    for  multiple  targets \cite{saha}.

The problem  related  to searching for multiple targets in reasonably fast  time  including  searching for the exceptional configurations  can be solved by choosing a   different  coin operator, $\mathcal{C}_{g}$ \cite{giriepjd}, which searches  for  the self-loops of the target vertices.  This operator works   with only the  discrete-time lackadaisical quantum walk \cite{wong3}.  It has  been observed that  $\mathcal{C}_{g}$ performs    better  compared  to the other known coin  operators previously studied in the literature.  It can search for a  single  target as well as multiple  target vertices with any configurations  with high success  probability.

This paper is arranged   in the following fashion:  A  discussion on the  quantum walk  search   on the  Johnson graph   is  presented  in Section    \ref{qw}.   Multi-target  search   on a complete graph is studied  in Section \ref{com},  on a triangular graph in Section \ref{tr} and on   $J(n,k\ge 3)$ graph in Section \ref{john}.  Finally,    we conclude in Section \ref{con} with a discussion.  

\section{QW search on Johnson graph} \label{qw}
In this section we  study  multi-target  spatial search  on  a  Johnson graph, $J(n,k)$ using discrete-time  coined quantum walk.  
Because  $J(n,k)$ is isomorphic to   $J(n,n-k)$,  we restrict ourself  to   $n \geq 2k$ case only.   
We represent $N$ vertices  and $d$ edges of the Johnson  graph as  the basis states  of  the Hilbert space of vertices   $\mathcal{H}_{vtx}$ and the  space of edges   $\mathcal{H}_{edg}$ respectively.   The initial state of the vertex space is the uniform superposition of all the $N$ basis states  $|m \rangle$   of the Hilbert space  $\mathcal{H}_{vtx}$:
\begin{eqnarray}
 |\psi_{vtx}\rangle =   \frac{1}{\sqrt{N}}
\sum_{m} |m \rangle \,.
 \label{in1}
\end{eqnarray}
Similarly the initial state for the coin space is the uniform superposition of all the edges associated with the vertex  $m$
\begin{eqnarray}
 |\psi_{edg}\rangle   =   \frac{1}{\sqrt{d}}  \sum_{m_n} |m_n\rangle \,,
 \label{in2}
\end{eqnarray}
where the summation is over all  the $d$ basis states  $|m_n\rangle$ of the Hilbert space  $\mathcal{H}_{edg}$.  Note that  $m$ in   $|m_n\rangle$ refers to the  vertex label  to which the edge  basis state belongs  and the suffix $n$ refers to the  nearest neighbor  vertex label the edge basis state  points  to.
Since discrete-time quantum walk evolves  in the  tensor product  space  $\mathcal{H} = \mathcal{H}_{vtx} \otimes \mathcal{H}_{edg}$,  the initial state for the quantum walk process is given by 
\begin{eqnarray}
 |\psi_{in}\rangle =  |\psi_{vtx}\rangle \otimes   |\psi_{edg}\rangle  =   \frac{1}{\sqrt{N d}} 
\sum_{m}  \sum_{m_n } |m \rangle  \otimes |m_n \rangle\,.
 \label{in3}
\end{eqnarray}
\begin{figure*}
\centering
\includegraphics[width=0.70\textwidth]{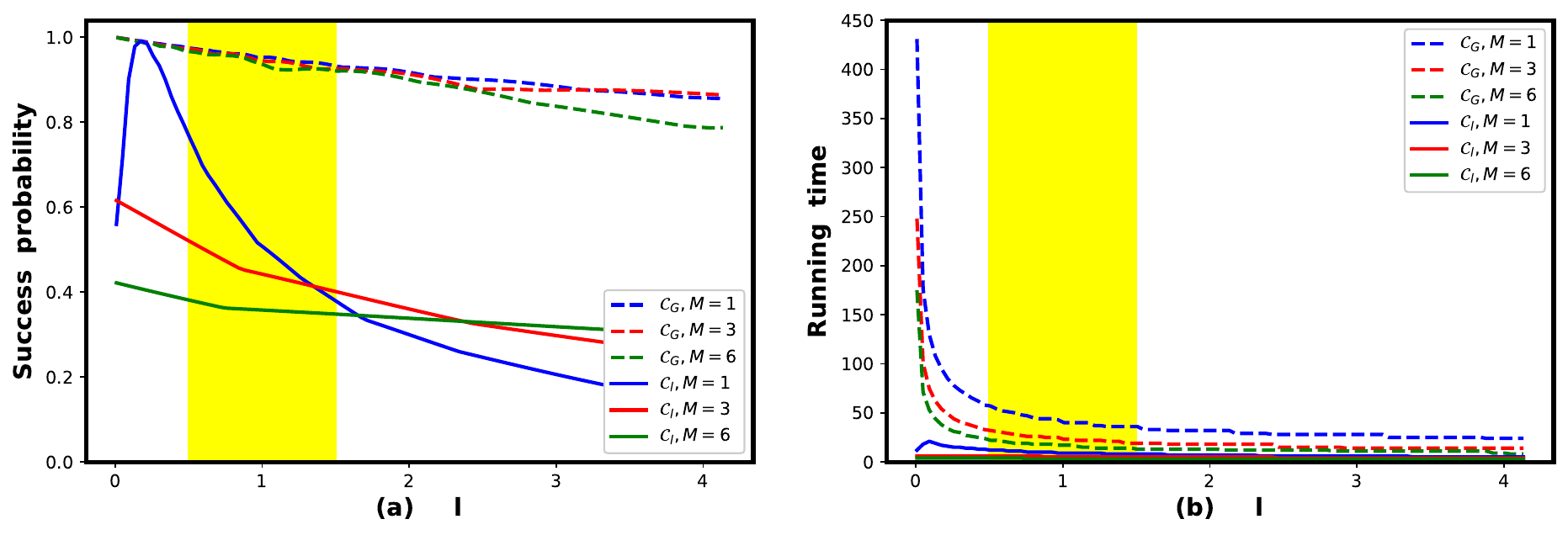}
\caption{\textbf{(a) Variation of success probability   and  (b) running time   as a function of the self-loop weight $l$  for a Johnson graph $J(10,3)$}}. 
\label{fig2}
\end{figure*}
The basis state  $|m \rangle  \otimes |m_n \rangle$,  belonging  to the tensor product space $\mathcal{H}$,   represents  state associated with the vertex $m$  with the edge pointing towards the vertex  $n$.  

For the lackadaisical quantum walk, as mentioned before,    we need to add a self-loop  of weight $l$ at each vertex of the graph, which creates an additional edge.  This is the quantum analog  of the  lazy random walk  in the classical regime. The self-loop allows  the corresponding  probability amplitude to stay put  \cite{wong1} at the same vertex, which helps to  accumulate  probability amplitude  at the target vertices in  quantum walk search. It has been observed that lackadaisical quantum walk  can  search  for target vertices on a two-dimensional periodic lattice including other graphs  with very high success probability without    any additional amplitude amplification technique, which is usually required  in case of  standard quantum walk  search (without self-loop)  by Grover or SKW coins.
The  Hilbert space of the edge,  $\mathcal{H}_{edg}$, becomes $d+1$-dimensional, because of one additional edge  of the self-loop. 
Initial state for the coin space at a vertex $m$ is then  given by 
\begin{eqnarray}
 |\psi_{edg}\rangle   =   \frac{1}{\sqrt{d +l}} 
\left[ \sum_{m_n \neq m_m}  |m_n \rangle  + \sqrt{l} |m_m \rangle \right]\,,
 \label{in4} 
\end{eqnarray}
and  the initial state for the quantum walk process is given by 
\begin{eqnarray} \nonumber 
 |\psi_{in}\rangle =   \frac{1}{\sqrt{N(d+l)}}  \times \hspace{4cm} \\
\sum_{m}  \left[ \sum_{m_n \neq m_m} |m \rangle  \otimes |m_n  \rangle +  \sqrt{l}|m \rangle  \otimes |m_m \rangle \right]\,.
 \label{in5}
\end{eqnarray}
Depending on whether standard  or lackadaisical quantum walk  is involved in the  search algorithm,  we  need to accordingly select  the initial state  between  eq. (\ref{in3}) and  eq. (\ref{in5}) respectively.  Then the evolution operation $\mathcal{U} = S \mathcal{C}$,  composed  of   modified coin operator  $\mathcal{C}$ followed by the  flip-flop shift  operator $S$, is applied to the initial state repeatedly until the target states are obtained with high success probability.  

Different types of coin operators exist, which are crucial for the quantum walk search.  Let us assume that  $M$ target vertices collectively   form  a set 
$\mathcal{T}_M$, which we have to find out by quantum walk search.   For simplicity of numerical evaluations, we chose first  $M$  vertices  from the list of  $N$ vertices  obtained from  sorted  tuples  $combinations(n, p)$  for  a  Johnson graph  $J(n, k)$. An example of the sorted order of vertices  of  the graph  $J(4, 2)$ in fig. \ref{fig1}(right)   is  $\{ 12, 13, 14, 23, 24, 34 \}$.

Then the  the modified coin   operator  $ \mathcal{C}_g$ of ref. \cite{giriepjd} acts as 
\begin{eqnarray} \nonumber
 \mathcal{C}_g |m \rangle  \otimes |m_n \rangle  = \hspace{4.5cm} \\
 \begin{cases}
 ~~C|m \rangle  \otimes |m_n \rangle & \text{if  $m  \notin  \mathcal{T}_M$}\\
 ~~C|m \rangle  \otimes |m_n \rangle &\text{if  $m  \in  \mathcal{T}_M$ and  $m_n \neq m_m$}\\
 - C| m \rangle  \otimes |m_n \rangle & \text{if  $m  \in  \mathcal{T}_M$ and  $m_n = m_m$}
\end{cases}  
\label{coing}           
\end{eqnarray}
where  $C= 2|\psi_{edg}\rangle \langle \psi_{edg}| - \mathbb{I}_{d+1\times d+1}$  is the Grover  diffusion operator.  Note the  difference of the above coin with the   coin operator  $\mathcal{C}_l$ of the lackadaisical quantum walk search
\begin{equation}
 \mathcal{C}_l |m \rangle  \otimes |m_n \rangle  =
    \begin{cases}
      ~~C|m \rangle  \otimes |m_n \rangle & \text{if  $m  \notin  \mathcal{T}_M$}\\
      - C|m \rangle  \otimes |m_n \rangle & \text{if  $m  \in  \mathcal{T}_M$}
      \end{cases}  
 \label{coinl}          
\end{equation}
\begin{figure*}
  \centering
     \includegraphics[width=0.70\textwidth]{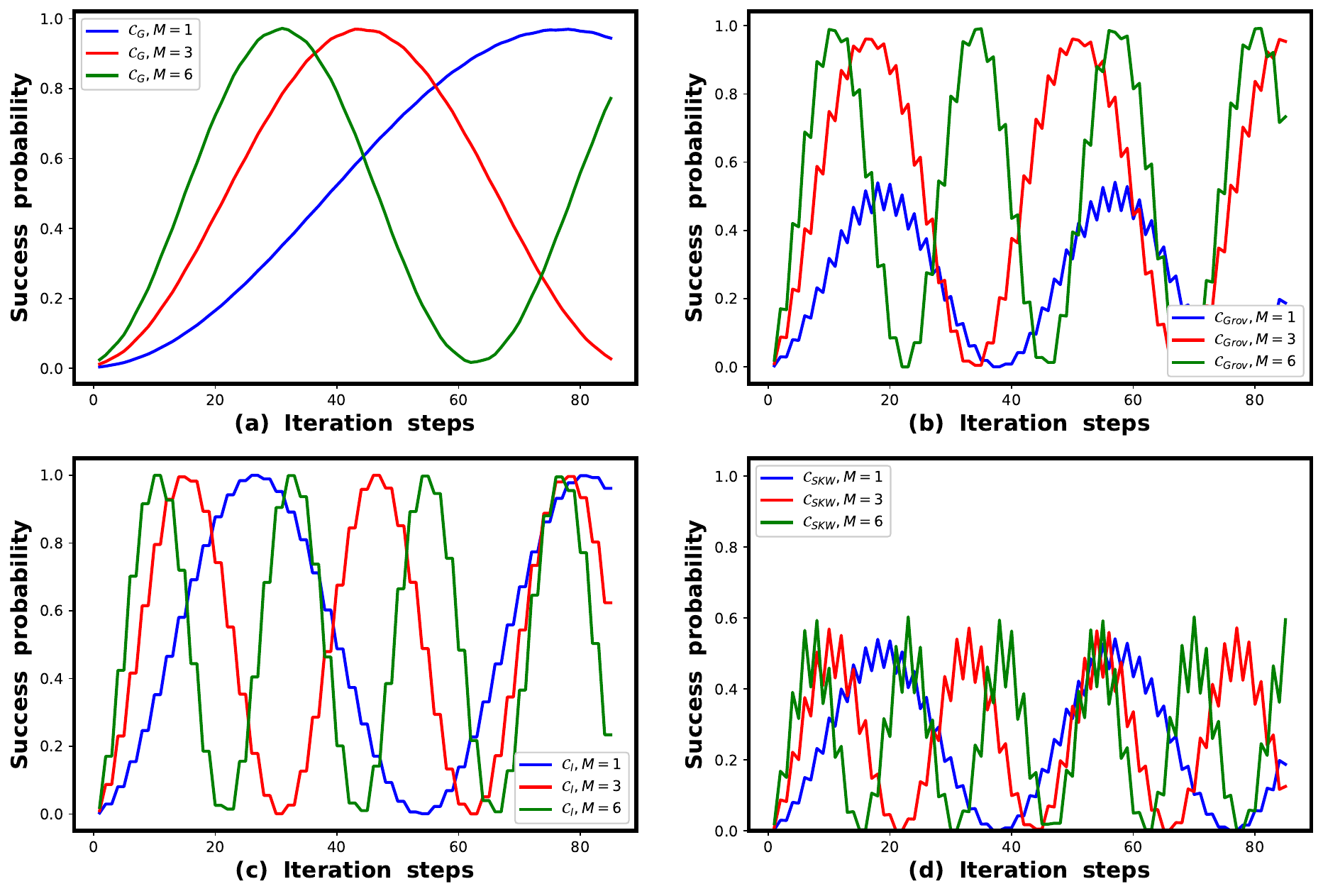}
          
       \caption{\textbf{Multi-target quantum walk search on a  complete graph with  $N=300$ vertices  and with  (a) $\mathcal{C}_{g}$, $l=10$ (b) $\mathcal{C}_{grov}$, (c) $\mathcal{C}_{l}$, $l=1$ and (d) $\mathcal{C}_{skw}$ coins for  $M=1$(blue), $3$(red), and $6$(green) targets.}}
\label{fig3}       
\end{figure*}
The above two coin operators are used  in  lackadaisical quantum walk with a self-loop.   The coin operator  $\mathcal{C}_g$    does   Grover search  on the edge space  of   the target  vertices with only the self-loop as the target, contrary to  other known coin operators, which basically search  for the target vertices, i.e., search for all the edge basis states of the target vertices. In effect,  $\mathcal{C}_g$  mostly   allows  inward flow of the  probability amplitude, with very small amount of  probability amplitude going out.

There are other two well known coin operators  $\mathcal{C}_{grov}$  and  $\mathcal{C}_{skw}$  for the  quantum walk search, which work on a graph with no self-loops.   Grover coin  $\mathcal{C}_{grov}$ acts as  eq. (\ref{coinl}) with the exception that now   $C = 2|\psi_{edg}\rangle \langle \psi_{edg}| - \mathbb{I}_{d\times d}$ is obtained from  eq. (\ref{in2}).  So,  $\mathcal{C}_{l}$ is the generalization of $\mathcal{C}_{grov}$  to lackadaisical quantum walk.   Finally  $\mathcal{C}_{skw}$ coin  is given by 
\begin{equation}
 \mathcal{C}_{skw} |m \rangle  \otimes |m_n \rangle  =
    \begin{cases}
      ~~C|m \rangle  \otimes |m_n \rangle & \text{if  $m  \notin  \mathcal{T}_M$}\\
      - I |m \rangle  \otimes |m_n \rangle & \text{if  $m  \in  \mathcal{T}_M$}
      \end{cases}  
 \label{coinskw}          
\end{equation}
Note that,  since only identity operator $I$ acts on the target vertex state   $\mathcal{C}_{skw}$  cannot  have nontrivial generalization  to lackadaisical quantum walk. 
The shift operator acts on the $Nd$  basis states of the combined Hilbert space  as 
\begin{eqnarray} 
S|m \rangle  \otimes |m_n \rangle  = | n \rangle  \otimes |n_m \rangle, ~~ m \ne n \,.
\label{shift}
\end{eqnarray}
In case of  $\mathcal{C}_{g}$  and $\mathcal{C}_{l}$, since  there are self-loops in addition to the regular edges, 
the  $N$  basis states  $| m \rangle  \otimes |m_m \rangle$  with  attached  self-loop acted by the shift operator,    remain  same   as   
\begin{eqnarray} 
S|m \rangle  \otimes |m_m \rangle  = | m \rangle  \otimes |m_m \rangle\,.
\label{shifts}
\end{eqnarray}
The action of the shift operator in eq. (\ref{shift})  can be better understood from its application on the complete graph $J(5,1)$  in fig. \ref{fig1}(left).  The vertices  $1, \cdots, 5$ are represented by the basis states  $|1 \rangle  \cdots,  |5 \rangle$   of the vertex space. Each vertex has four associated edge basis states.  For example, the vertex state  $|1 \rangle$ has  four associated edge basis states   $|1_2 \rangle$, $|1_3 \rangle$, $|1_4 \rangle$ and $|1_5 \rangle$, which are shown  as  the four red arrows from left to right respectively.    Note that  the suffixes $2, \cdots, 5$ of the four edge basis states    represent  the vertices  they point to.  The action of the  shift operator $S$  on the tensor product state  $|1\rangle \otimes  |1_2\rangle$  is as follows:  $S|1 \rangle  \otimes |1_2 \rangle  = | 2 \rangle  \otimes |2_1 \rangle$.  Similarly,   $S$ acts on all of the twenty basis states  of the tensor product space  $\mathcal{H}$.
In lackadaisical quantum walk, additionally   $S$ acts on the vertex $1$ with the self-loop in fig. \ref{fig1}(left)  as $S|1 \rangle  \otimes |1_1 \rangle  = | 1 \rangle  \otimes |1_1 \rangle$. Note that in fig. \ref{fig1} only graph without  any self-loop has been presented. But  when  considering lackadaisical quantum walk, we have to keep in mind that there is a self-loop at each vertex of the graph. 

The total success probability after $t$ time steps  to find one of the $M$ targets, belonging to the set  $\mathcal{T}_M$,  is given by 
\begin{eqnarray} 
p_{s} = \sum_{m \in \mathcal{T}_M} |\langle m  | \mathcal{U}^t |\psi_{in} \rangle|^2\,.
 \label{psucc}
 \end{eqnarray}
\begin{figure*}
  \centering
     \includegraphics[width=0.70\textwidth]{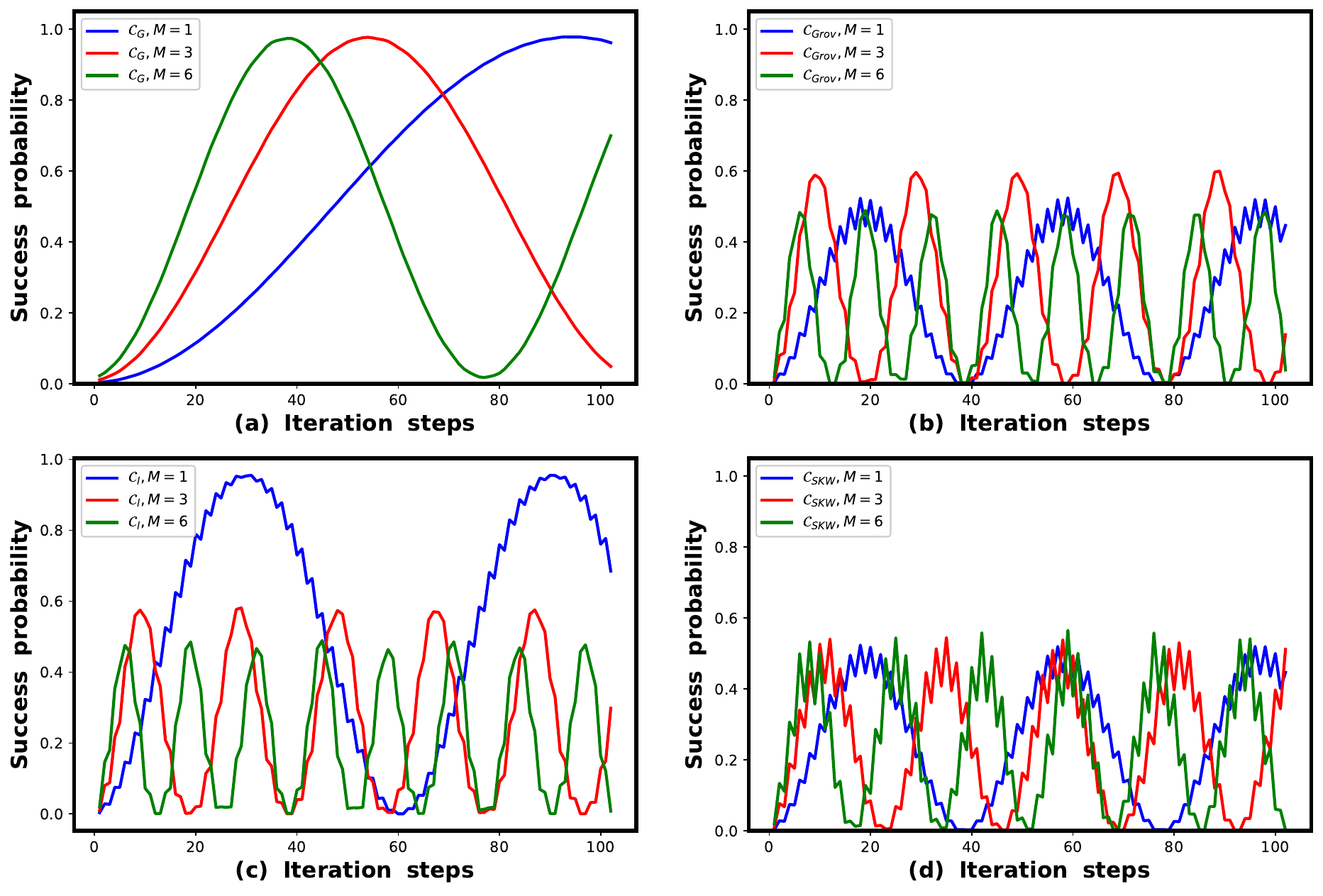}
          
       \caption{\textbf{Multi-target quantum walk search on a  $25$-triangular  graph $J(25,2)$  with   $N=300$ vertices and with (a) $\mathcal{C}_{g}$, $l=1$ (b) $\mathcal{C}_{grov}$, (c) $\mathcal{C}_{l}$, $l=0.1$ and (d) $\mathcal{C}_{skw}$ coins for  $M=1$(blue), $3$(red), and $6$(green) targets.}}
\label{fig4}
\end{figure*}
In  case of  search with  the  lackadaisical quantum walk, we need to find an optimal value of the self-loop parameter $l$, so that the success probability is maximized.     Two of the coin operators, $\mathcal{C}_G$ and $\mathcal{C}_l$, discussed in this article are based on the  lackadaisical quantum walk.
In order to understand the role of the self-loop on quantum walk search on  the Johnson graph $J(10,3)$, in figs.  \ref{fig2}(a)-(b), we plot success probability and running time  as a function of the self-loop weight $l$. We can see around the yellow strip region the success probability is very high and at the same time the running time saturates. So, we can fix  a value for  the self-loop in the yellow strip region and  run the evolution operator to obtain the final success probability.   Observe from   fig.  \ref{fig2}(a) that   the success probability achieved  by $\mathcal{C}_{g}$ coin is much higher than  $\mathcal{C}_{l}$. From the next section onward we will see that the success probability obtained by   $\mathcal{C}_{g}$  is always higher than  that obtained by  $\mathcal{C}_{l}$ and others coins studied in this article.  

Usually $l$ depends on the graph, degree of the graph,  number of targets $M$ etc. Therefore,  we need to  obtain an optimum value of the self-loop  by fixing the parameters of the  Johnson graph $J(n, k)$ and the number of target states $M$ and then running the exercise of fig. \ref{fig2}.  However,  to save time,  we choose a  fixed value of the self-loop parameter, that  provides a reasonably high success probability, to study our multi-target quantum search.   

From the next section onward  we study multi-target search of the Johnson graph $J(n, k)$ for different $k$.  Since,  $k=1$ and $2$ correspond to the  special graphs, known as  the  complete and triangular  graph respectively,   
we devote  the next two sections  to study these two graphs first. Then   we discuss   $J(n, k \ge 3)$. 
\section{Complete graph} \label{com}
A complete graph  $K_n$ can be obtained by setting  $k=1$ in  the Johnson graph  $J(n,k)$.  It has $N=n$ vertices and each vertex has $d= N-1$ edges.   It is one of the earliest graph on which quantum walk  has been used to search  for a target vertex.  It can  been shown that the quantum walk search with Grover coin  on the complete graph with a self-loop at each vertices  is equivalent to the Grover search on  both the vertex space and the coin  space \cite{portugal}.  Single target search on complete graph with lackadaisical quantum walk  with equal wight self-loop at each vertex  has been  studied in ref.   \cite{wong2}, which shows that for $l=1$ success probability is $\mathcal{O}(1)$.    In ref.   \cite{rapo}  symmetry of the  graph   is broken by  using different  self-loop weight in  each vertex, which shows that only the wight of the self-loop at the marked vertex matters.

In this  section, we consider  multi-vertex search on this graph by  lackadaisical quantum walk with $\mathcal{C}_{g}$ coin and compare the result with other coins.  The initial state of the $N$-dimensional  vertex space  $\mathcal{H}_V= C^N$  is given by  the uniform superposition of the basis states 
\begin{eqnarray}
 |\psi_{vtx}\rangle =   \frac{1}{\sqrt{N}}
\sum_{m=1}^{N} |m \rangle \,.
 \label{in1com}
\end{eqnarray}
Although, complete graph has $d=N-1$ degree, the coin space becomes $N$-dimensional after we add one self-loop at each vertex of the graph.  Then the initial state for the  coin space  $\mathcal{H}_C = C^N$ at a vertex $m$ is given by  
\begin{eqnarray}
|\psi_{edg}\rangle   =   \frac{1}{\sqrt{N-1 +l}} 
\left[ \sum_{m_n \neq m_m}  |m_n \rangle  + \sqrt{l} |m_m \rangle \right].
 \label{in4com} 
\end{eqnarray}
We have studied quantum search to find $M=1, 3$ and $6$  target vertices  using four different coins and result of success probabilities  to find these target vertices  are plotted in fig. \ref{fig3}.  

{\it $\mathcal{C}_{g}~ coin$}:  Success probabilities for $M=1, 3$ and $6$,  represented by  blue, red and green  curves respectively,  are  plotted in fig. \ref{fig3}(a).  We have fixed the self-loop weight at $l=10$, which provides very high success probabilities in all the  cases.  

{\it $\mathcal{C}_{grov}~ coin$}:  As expected, running  time  to  search  a single target,   $M=1$  and success probability   as  represented by the blue curve in fig. \ref{fig3}(b)  agree  with  the analytical value  $\pi\sqrt{N}/(2\sqrt{2})$  and    $0.5$ \cite{tana} respectively.  Although  success probabilities   for $M=3$ and $6$,  represented  by  red and  green curves respectively,    are  very high. 

{\it $\mathcal{C}_{l}~ coin$}: In lackadaisical quantum walk with associated Grover coin  we  fix the self-loop weight at  $l=1$. Success probability for $M= 1, 3$ and $6$, represented by blue, red and green curves  in figs. \ref{fig3}(c) respectively,  are all very high. 

{\it $\mathcal{C}_{skw}~ coin$}: With  this coin  success probabilities for  $M= 1, 3$ and $6$ are  $\sim 0.5$, represented by blue, red and green curves  in figs. \ref{fig3}(d) respectively,   Of course, we can increase the success probability further by  using amplitude amplification technique \cite{bras}.

\section{Triangular graph} \label{tr}
Triangular  graph  $T_n$ can be obtained by setting  $k=2$ in  the Johnson graph  $J(n,k)$.  It has $N=n(n-1)/2$ vertices and each vertex has $d= 2(n-2)$ edges.   It is  an example of a strongly  regular graph $G(N, d, \lambda, \mu)$ \cite{rapo} with the values of the four parameters being   $N=n(n-1)/2$,  $d= 2(n-2)$, $\lambda = n-2$, and $\mu =4$. This system has been studied in  ref. \cite{rapo}  to search for a single target vertex. It shows that for a specific value of the self-loop parameter the success probability of the lackadaisical quantum walk search can achieve $\mathcal{O}(1)$.  

Bellow we report our  result to search multiple targets on a  $25$-triangular/$J(25,2)$ graph  using four coin operators.  
$J(25,2)$ has $N=300$ vertices and $d= 46$ nearest neighbor vertices at each vertex. 

{\it $\mathcal{C}_{g}~ coin$}:  Success probabilities for $M=1, 3$ and $6$ target vertices,  represented by  blue, red and green  curves respectively,  are  plotted in fig. \ref{fig4}(a).  We have fixed the self-loop weight at $l=1$, which provides very high success probabilities in all the three   cases.  

{\it $\mathcal{C}_{grov}~ coin$}:  As expected, running  time  to  search  a single target,   $M=1$  and success probability   as  represented by the blue curve in fig. \ref{fig4}(b)  agree  with  the analytical value  $\pi\sqrt{N}/(2\sqrt{2})$  and    $0.5$ \cite{tana} respectively.  However, contrary to the complete graph case, now  success probabilities   for $M=3$ and $6$ target vertices,  represented  by  red and  green curves respectively,   are $\sim 0.5$. 

{\it $\mathcal{C}_{l}~ coin$}: In lackadaisical quantum walk with associated Grover coin  we  fix the self-loop weight at  $l=0.1$. Success probability for $M= 1$  target vertex, represented by blue curve  in figs. \ref{fig4}(c),  is  very high.  However, success probabilities for  $M= 3$ and $6$ target vertices are $\sim 0.5$. 

{\it $\mathcal{C}_{skw}~ coin$}: With  this coin,   success probabilities for  $M= 1, 3$ and $6$ target vertices are  $\sim 0.5$, represented  by blue, red and green curves  in figs. \ref{fig4}(d) respectively.  Of course, we can increase the success probability further by  using amplitude amplification technique. 
\begin{figure*}
  \centering
     \includegraphics[width=0.80\textwidth]{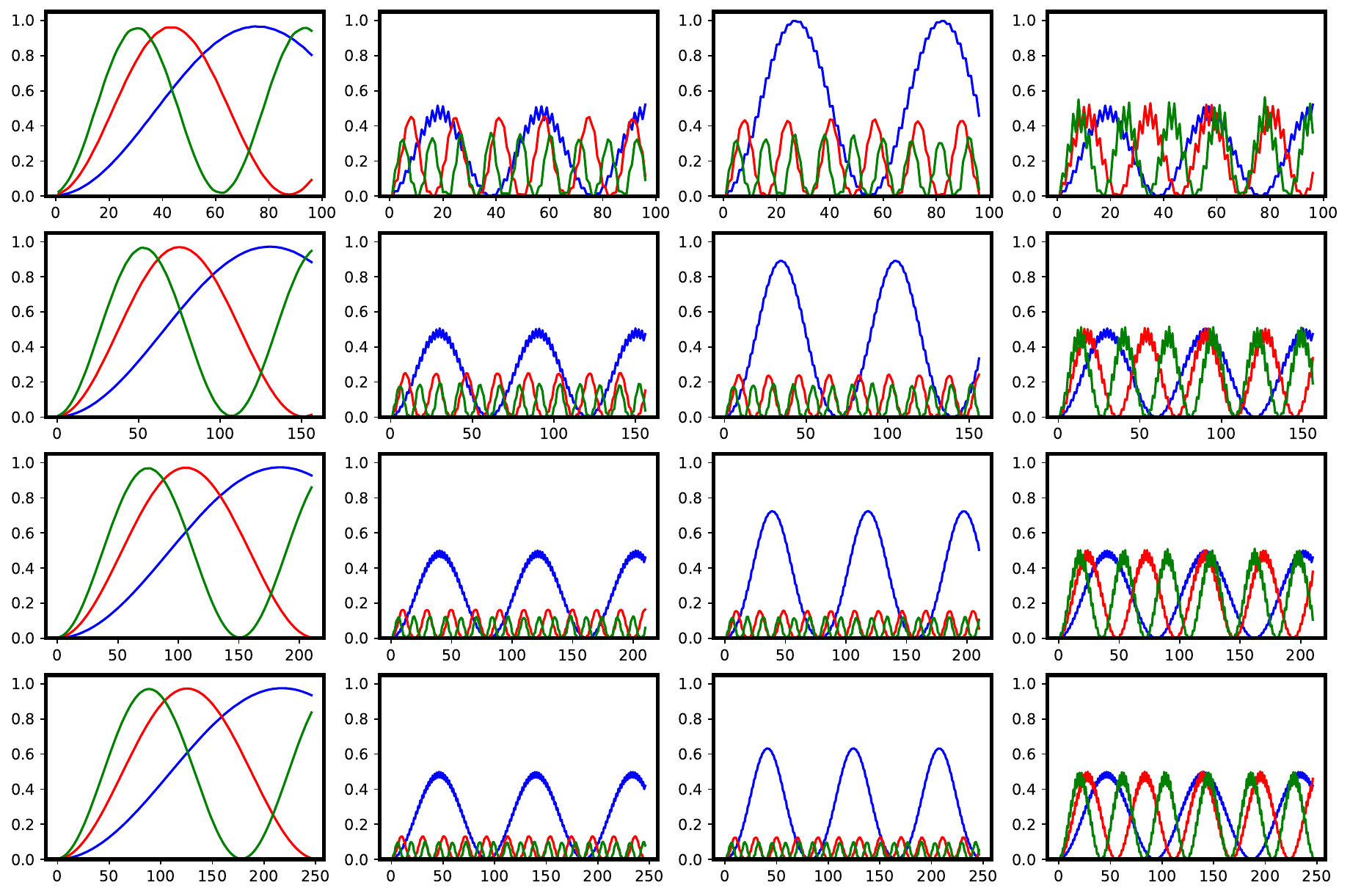}
          
       \caption{\textbf{Multi-target quantum walk search on Johnson   graph for $M=1$(blue), $3$(red), and $6$(green) targets.  Top to bottom rows  correspond  to  the Johnson graphs $J(13,3)$,  $J(13,4)$,   $J(13,5)$  and   $J(13,6)$ respectively.  Left to right columns correspond to  $\mathcal{C}_{g}$, $l=1.0$,    $\mathcal{C}_{grov}$,   $\mathcal{C}_{l}$, $l=0.1$  and   $\mathcal{C}_{skw}$  coins respectively.}}
\label{fig5}
\end{figure*}
\section{Johnson  graph $J(n,k\ge 3)$} \label{john}
In this section,  we consider  multi-target search on Johnson graphs  for $k \ge 3$. Specifically,  for $k=3$, it is  a  tetrahedral graph  with $N= n(n-1)(n-2)/6$  vertices and each vertex having $d= 3(n-3)$  nearest-neighbor  vertices.  It has been shown in ref. \cite{wongjpa} that, like complete and triangular graph,  this graph also allows fast quantum walk search  for a single target with high success probability.

Bellow we report our  result to search multiple targets on   $J(13, 3)$, $J(13, 4)$, $J(13, 5)$, and $J(13, 6)$ Johnson  graphs  using four coin operators.   From top to bottom, first row corresponds to  $J(13, 3)$ with $N = 286$ vertices and $d = 30$ degree, second  row corresponds to  $J(13, 4)$ with $N = 715$ vertices and $d = 36$ degree, third  row corresponds to  $J(13, 5)$ with $N = 1287$ vertices and $d = 40$ degree, and  fourth row corresponds to  $J(13, 6)$ with $N = 1716$ vertices and $d = 42$ degree respectively.  Left to right columns correspond to  $\mathcal{C}_{g}$, $\mathcal{C}_{grov}$, $\mathcal{C}_{l}$, and $\mathcal{C}_{skw}$  coins respectively.

{\it $\mathcal{C}_{g}~ coin$}:  Success probabilities for $M=1, 3$ and $6$,  represented by  blue, red and green  curves respectively,  are  plotted in first column of  fig. \ref{fig5}.  We have fixed the self-loop weight at $l=1$, which provides very high success probabilities in all the  cases for all the four Johnson graphs.  

{\it $\mathcal{C}_{grov}~ coin$}:  As expected, running  time  to  search  a single target,   $M=1$  and success probability   as  represented by the blue curve in second column of  fig. \ref{fig5}  agree  with  the analytical value  $\pi\sqrt{N}/(2\sqrt{2})$  and    $0.5$ \cite{tana} respectively.  However, contrary to the complete graph case, now  the success probabilities   for $M=3$ and $6$ target vertices,  represented  by  red and  green curves respectively,   gradually decrease  bellow $\sim 0.5$ as $k \ge 3$ increases.

{\it $\mathcal{C}_{l}~ coin$}: In lackadaisical quantum walk with associated Grover coin  we  fix the self-loop weight at  $l=1$ for all the four Johnson graphs. Success probabilities  for $M= 1$, represented by blue curves  in third column of figs. \ref{fig5}  are  high. Note that the gradual decrease of success probability can be improved by choosing optimum self-loop weight for  that particular Johnson graph.   However,  it is observed that the success probabilities   for $M=3$ and $6$ target vertices,  represented  by  red and  green curves respectively,   gradually decrease  bellow $\sim  0.5$ as $k \ge 3$ increases---similar to the case observed by  $\mathcal{C}_{grov}$ coin.

{\it $\mathcal{C}_{skw}~ coin$}: Using   this coin  success probabilities for  $M= 1, 3$ and $6$  target vertices are  $\sim 0.5$, represented  by blue, red and green curves  in fourth column of  fig. \ref{fig5}  respectively.  Of course, we can increase the success probability further by  using amplitude amplification technique.

\section{Conclusions} \label{con}
Discrete-time quantum walk  is a widely used  tool to perform spatial search for  target vertices on  several  graphs.   Although  single target search has been successfully  implemented  on various graphs,  searching for multiple targets on graphs  comes with challenges. For example, certain  types of configurations of target vertices, known as  exceptional configurations,  are hard to search in both  standard and lackadaisical quantum walk.  Also, different  coin operators, such as $\mathcal{C}_{g}$, $\mathcal{C}_{grov}$, $\mathcal{C}_{l}$, and $\mathcal{C}_{skw}$ coins,   behave differently  while  searching  of target vertices on a graph.   

In this article we explore  multi-target spatial search on Johnson graphs $J(n, k)$ for different values of the parameter $k$.  We numerically analyze  and compare  the performances  of four different coin operators to search multiple target vertices on Johnson graphs.  We observe that  the $\mathcal{C}_{g}$ coin can search multiple targets on any Johnson graphs $J(n, k)$ discussed in this article with very high success probability. 
In the case of the $\mathcal{C}_{grov}$ coin, running  time  to  search  $M=1$  target  and  its corresponding success probability   agree  with  the analytical value  $\pi\sqrt{N}/(2\sqrt{2})$  and    $0.5$   respectively for the the Johnson graphs discussed in this article.   Although  success probabilities   for $M=3$ and $6$,  represented  by  red and  green curves respectively,  become   very high for  $k = 1$(complete graph), it gradually decrease  below  $\sim 0.5$ as $k$ increases.    For $M=3$ and $6$ targets, a  similar behavior is  observed for  the  $\mathcal{C}_{l}$ coin aa well. However, an $M = 1$ target  can be searched by  the  $\mathcal{C}_{l}$ coin with high success probability for all the Johnson graphs. 
In the case of the $\mathcal{C}_{skw}$ coin, success probability to search  $M =1, 3$, and $6$  is  $0.5$ for all the Johnson graphs.   Observations  from our numerical analysis suggest that among the four coins discussed in this article  the $\mathcal{C}_{g}$  coin can search multiple targets with very high success probability on all the Johnson graphs. 

As a future work, it would be interesting, though challenging, to analytically  calculate  the time complexity and success probability for multi-target  search by quantum walk using the coin operators  discussed  in this article. 

\vspace{1cm}

Data availability Statement:  The data  generated during and/or analyzed   during the current study is  included in the article.
\vspace{0.5cm}

Conflict of interest: The authors have no competing interests to declare that are relevant to the content of this article. 


\end{document}